\documentclass[prb,onecolumn]{revtex4-1}

\usepackage{graphicx}
\large
\usepackage{amsmath}
\usepackage{float}
\usepackage{gensymb}

\large

\begin{document}


\title{Electrically pumped WSe$_2$-based light-emitting van der Waals heterostructures embedded in monolithic dielectric microcavities}

\author{O. Del Pozo-Zamudio$^{1,7}$}
\author{A. Genco$^{1\ast}$}
\author{S. Schwarz$^{1}$}
\author{F. Withers$^{2,3}$}
\author{P. M. Walker$^1$}
\author{T. Godde$^1$}
\author{R. C. Schofield$^1$}
\author{A. P. Rooney$^4$}
\author{E. Prestat$^4$}
\author{K. Watanabe$^5$}
\author{T. Taniguchi$^5$}
\author{C. Clark$^6$}
\author{S. J. Haigh$^4$}
\author{D. N. Krizhanovskii$^1$}
\author{K. S. Novoselov$^{2,8,9}$}
\author{A. I. Tartakovskii$^{1\ast\ast}$}

\affiliation{$^1$Department of Physics and Astronomy, University of Sheffield, Sheffield S3 7RH, UK}

\affiliation{$^2$School of Physics and Astronomy, University of Manchester, Manchester M13 9PL, UK}

\affiliation{$^3$Centre for Graphene Science, CEMPS, University of Exeter, Exeter, EX4 4QF, UK}

\affiliation{$^4$School of Materials, University of Manchester, Manchester M13 9PL, UK}

\affiliation{$^5$National Institute for Materials Science, 1-1 Namiki, Tsukuba 305-0044, Japan}

\affiliation{$^6$Helia Photonics, Livingston EH54 7EJ, UK}

\affiliation{$^7$Instituto de Investigacion en Comunicacion Optica, Universidad Autónoma de San Luis Potosí, San Luis Potosi, San Luis Potosí, MEXICO}

\affiliation{$^8$Centre for Advanced 2D Materials, National University of Singapore, 117546 Singapore}

\affiliation{$^9$Chongqing 2D Materials Institute, Liangjiang New Area, Chongqing, 400714, China}

\date{\today}

\author{$\ast$ a.genco@sheffield.ac.uk; $\ast\ast$ a.tartakovskii@sheffield.ac.uk}

\begin{abstract}
Vertical stacking of atomically thin layered materials opens new possibilities for the fabrication of heterostructures with favorable optoelectronic properties. The combination of graphene, hexagonal boron nitride and semiconducting transition metal dichalcogenides allows fabrication of electroluminescence (EL) devices, compatible with a wide range of substrates. Here, we demonstrate a full integration of an electroluminescent van der Waals heterostructure in a monolithic optical microcavity made of two high reflectivity dielectric distributed Bragg reflectors (DBRs). Owing to the presence of graphene and hexagonal boron nitride protecting the WSe$_2$ during the top mirror deposition, we fully preserve the optoelectronic behaviour of the device. Two bright cavity modes appear in the EL spectrum featuring Q-factors of 250 and 580 respectively: the first is attributed directly to the monolayer area, while the second is ascribed to the portion of emission guided outside the WSe$_2$ island. By embedding the EL device inside the microcavity structure, a significant modification of the directionality of the emitted light is achieved, with the peak intensity increasing by nearly two orders of magnitude at the angle of the maximum emission compared with the same EL device without the top DBR. Furthermore, the coupling of the WSe$_2$ EL to the cavity mode with a dispersion allows a tuning of the peak emission wavelength exceeding 35 nm (80 meV) by varying the angle at which the EL is observed from the microcavity. This work provides a route for the development of compact vertical-cavity surface-emitting devices based on van der Waals heterostructures. 
\end{abstract}

\maketitle

\section{Introduction}

The fabrication of heterostructures made of atomically thin layers of two-dimensional (2D) materials enables the development of novel optoelectronic devices with a wide range of properties dependent on the combinations of the layers in the stack\cite{Dean2010,NovoselovPNAS2005,Geim2013}. Such layers are held in place by van der Waals forces which allow a variety of 2D materials with different lattice constants and crystal axes orientations to be combined {\it ad hoc}\cite{Geim2013}. Semiconducting transition metal dichalcogenide (TMD) monolayers play a fundamental role in such devices, owing to their attractive optical properties\cite{Britnell2013,WangNatNano2012,XuNatPhys2014}. TMDs experience an indirect-to-direct band-gap transition as the material is thinned to a monolayer\cite{MakPRL2010}, and exhibit excitons stable at room temperature \cite{Ugeda2014,Berkelbach2013}, with high binding energies and large oscillator strengths \cite{Mak2016}. These properties have been already utilized in electroluminescence (EL) devices \cite{Pospischil2014,Ross2014,Withers2015,Withers2015NanoLett,EdaAdvMat2018}, most commonly employing a design based on atomically thin layers of TMDs as the light emitting material, clad by insulating hexagonal boron nitride and graphene electrodes \cite{Withers2015,Withers2015NanoLett}.

The optical properties of an electroluminescent device can be additionally tailored by inserting the structures in an optical microcavity, thus strongly modifying the spectral and directional properties of its emission as well as the optical gain \cite{SchubertScience1994}. This approach is widely used in opto-electronics, and is fundamental for compact vertical cavity surface emitting lasers (VCSELs), where an active medium, an electrically pumped semiconductor quantum well, is placed between two high reflectivity mirrors\cite{Soda1979,Koyama1989}. A narrow spectral linewidth and a high degree of  angular directionality of an electroluminescent microcavity are also important for telecommunications applications\cite{Michalzik2012}. So far the coupling of TMD monolayers to optical cavities of various designs has been exploited to study photoluminescence (PL) enhancement and the Purcell effect\cite{Wu2014,Gan2013,Schwarz2014}, to obtain lasing\cite{Wu2015,Salehzadeh2015,Ye2015,Shang2017}, and to reach the regime of the strong light-matter interaction\cite{Liu2015,Dufferwiel2015,Lundt2016,Sidler2017,SchneiderReview2018}. A first demonstration of a TMD monolayer-based electroluminescent device coupled to an optical resonator was shown by employing a mechanically transferred L3 photonic crystal cavity (PC) made of GaP ontop of a graphene/hBN/WSe$_2$/hBN/graphene LED. A Q-factor below 100 and a 4-times increase of EL peak intensity compared to the same uncoupled device have been reported for this structure\cite{Liu2016}. Very recently an electroluminescent planar microcavity made of a stack of several WS$_2$ layers sandwiched between a distributed Bragg reflector (DBR) and a silver mirror has been reported, showing evidences of strong light-matter coupling at room temperature. Nevertheless, due to the strong optical dissipation of the metal mirror, a bare cavity mode linewidth of 24 meV has been measured leading to a relatively low Q value of about 80\cite{Gu2019}.

Here, we demonstrate a novel electroluminescent microcavity device embedding a van der Waals heterostructure of the form hBN/graphene/hBN/WSe$_2$/hBN/graphene between two high reflectivity dielectric DBRs. In order to show the full functionality of the devices, we directly compare the EL behaviour of the heterostructures placed on a DBR (a 'half-cavity' configuration), with those enclosed in the full monolithic cavity, with the top mirror deposited after fabrication of the EL device (a 'full-cavity' configuration). No degradation of the optical or electrical properties after the deposition of the top DBR has been observed, which we explain by the protective role of the graphene and hBN layers. We show that EL excites several microcavity modes arising from to the finite (and relatively small) dimensions of the heterostructure. The two main modes are attributed respectively to areas within the monolayer edges and to the portion of emission guided outside the WSe$_2$ island. Owing to the use of lossless dielectric mirrors, the latter showed a narrow linewidth of 1.3 nm, resulting in a remarkably high Q factor of 580. The cavity mode directly ascribed to the monolayer emitting region is also about one order of magnitude narrower than the bare WSe$_2$ EL emission spectrum, featuring a Q-factor of about 250, higher than the previously reported electroluminescent microcavity devices employing TMDs. We show strong changes in the angular distribution of the light emitted by the monolithic cavity resulting in a highly directional EL. We observe an EL peak intensity enhancement of about two orders of magnitude at the angle of the maximum emission compared with the devices in the half-cavity configuration. The microcavity EL peak can be tuned by up to 35 nm (or for 80 meV from 1.65 eV to 1.73 eV) by changing the angle of collection of the emitted light. By varying the operation temperature of the device, the angle at which the maximum intensity is observed can also be tuned. This work shows that the fabrication of monolithic cavities comprising atomically thin van der Waals electroluminescent structures is feasible, opening new avenues for studies of light-matter interaction in 2D materials under electrical excitation and for the development of novel TMD-based optoelectronic devices.

\section{Results and discussion}

\begin{figure*}
\includegraphics[width=.95\textwidth]{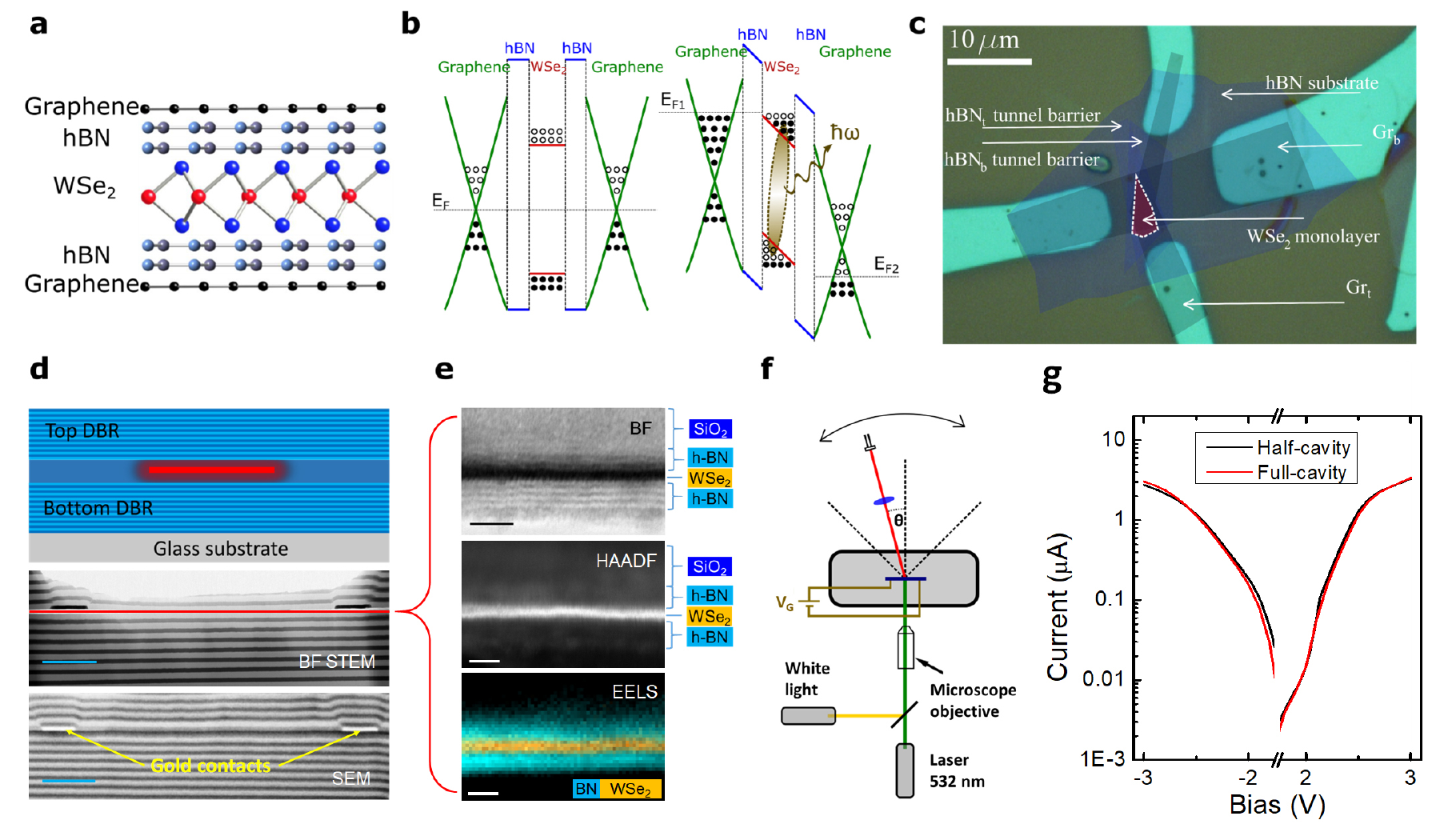}
\caption{\label{fig1} {\bf WSe$_2$ EL devices embedded in monolithic microcavities.} (a) Schematic diagram of the heterostructure constituting the EL device. (b) Electronic bandstructure of the EL device at zero bias voltage (left) and at the bias corresponding to the EL onset (right). (c) Optical image of the EL device prior to deposition of the top DBR. The area in which EL is observed (purple) is marked with the dashed line. All constituting layers of the van der Waals structure are labeled. (d) The top panel shows a schematic of the monolithic microcavity with an EL device embedded between the two DBRs. The lower panels show a bright field scanning transmission electron microscopy (BF STEM) image and a scanning electron microscopy (SEM) image of the cross-section one of the monolithic microcavity EL devices used in this study. Alternating dielectrics in the DBRs are visible as well as the layer distortions due to the gold contacts. The position of the EL device is marked by the red line. Scale bars correspond to 1 $\mu$m. (e) High resolution images of the EL device, measured at the same position. From top to bottom: STEM BF; STEM high angle annular dark field (HAADF); electron energy loss spectroscopy (EELS) chemical map highlighting positions of nitrogen and selenium atoms. Scale bars correspond to 2 nm. (f) Schematic of the optical set-up used for the angle-resolved measurements (see text for details). (g) I-V plot before (black lines) and after (red lines) the DBR deposition showing absence of degradation of the EL device.}
\end{figure*}

We use Gr/hBN/TMD/hBN/Gr heterostructures for our EL devices (Gr denotes single-layer graphene). Each device is assembled on a thin hBN layer placed on the bottom DBR. Graphene layers are used as highly transparent electrodes, which allow electric carrier injection through the hBN tunnel barriers. WSe$_2$ monolayer is the light-emitting material forming a quantum-well-like structure between the hBN barriers. A schematic of the heterostructure is shown in Fig.\ref{fig1}(a). The band-structure of the device in Fig.\ref{fig1}(b) is shown for zero bias between the graphene contacts (left) and at the EL onset voltage (right). As the bias voltage is increased, the Dirac points of the graphene contacts shift towards the band-edges of the WSe$_2$ monolayer. At a voltage where the quasi-Fermi level from one graphene contact exceeds the minimum of the conduction band of the TMD, electron injection into the active region occurs. By increasing the bias further, the quasi-Fermi level of the second graphene contact shifts below the maximum of the valence band which leads to hole injection.  The electrons and holes tunneling through the top and bottom barriers, respectively, form excitons in the WSe$_2$ film that recombine and give rise to photon emission\cite{Withers2015,Withers2015NanoLett}. In the devices investigated in our work, the EL onset is usually observed at the bias $V_b \approx$2.2 V \cite{Withers2015NanoLett}. An optical image of the device before deposition of the top dielectric mirror is shown in Fig.\ref{fig1}(c). The dashed lines highlight the area where EL is observed, corresponding to the position where all individual atomically thin layers constituting the heterostructure overlap.

We measure the same EL devices in half- and full-cavity configurations, by firstly characterizing the structures placed on a planar DBR before the top DBR is deposited using an electron beam evaporation technique. The schematic of the EL device, embedded in the microcavity is shown at the top of Fig.\ref{fig1}(d). The light-emitting device is transferred onto a dielectric mirror, a DBR consisting of 10 pairs of quarter-wavelength Nb$_2$O$_5$/SiO$_2$ layers designed for a central wavelength of 750 nm. We used Nb$_2$O$_5$ and SiO$_2$ layers with thicknesses of 127 and 95 nm and refractive indexes of 1.45 and 2.09, respectively. The mirror, grown on silica at 120 $\degree$C using an ion-assisted electron beam deposition, has a high reflectivity of more than 99\% at this wavelength and also provides a broad stop-band of 200 nm (see Supporting Information). An identical DBR with an inverse structure (SiO$_2$/Nb$_2$O$_5$) was fabricated on top of the EL device. However, it was made at a lower temperature of 60 $\degree$C and without ion-assisted deposition for the first 20 nm. The resultant $\lambda/2$ microcavity has the EL device situated in the center at the electric field antinode. Cross-sectional bright field scanning transmission electron microscopy (BF STEM) and scanning electron microscopy (SEM) images obtained on one of the devices embedded in the microcavity are shown in the bottom part of Fig.\ref{fig1}(d). The alternating layers of Nb$_2$O$_5$ and SiO$_2$ are clearly visible. It is also observed that the thickness of the gold contacts at the sides of the device is comparable with those of the $\lambda/4$ layers, which leads to  distortion of the DBR stack at the device edges (see Supporting Information for more detailed images). The position of the EL device is marked with the red line in Fig.\ref{fig1}(d). Fig.\ref{fig1}(e) displays high magnification cross-sectional STEM images (in Bright Field, BF, and in High Angle Annular Dark Field, HAADF) clearly showing the WSe$_2$ monolayer encapsulated in few layers of hBN. The WSe$_2$ is the dark (bright) layer in the BF (HAADF) image. The chemical map obtained using the electron energy loss spectroscopy (EELS) shown in Fig.\ref{fig1}(e) confirms that the WSe$_2$ monolayer is enclosed in hBN. Principal component analysis was used to improve the signal to noise ratio in the spectra comprising the EELS map. Further images and maps are given in the Supporting Information. 

The half-cavity devices were measured in a micro-electroluminescence (micro-EL) set-up where light is collected through a lens having numerical aperture, NA, of 0.55. A schematic of another optical set-up, used for the angle-resolved measurements on the  full-cavity devices, is shown in Fig.\ref{fig1}(f). The EL device is placed in a liquid-helium flow cryostat that allows experiments at temperatures from 10 to 300 K. A goniometer is used to rotate the signal collection path of the set-up in order to measure light emitted by, or transmitted through, the sample at different angles $\theta$, ranging from 0$\degree$ to $\pm$30$\degree$. In this configuration it is possible to carry out micro-transmission measurements where the excitation with  white light is focused on the sample through a microscope objective from one side of the sample, and the signal is measured from the other side with a required angular resolution. We also use this set-up to measure EL with high angular resolution. In the measurements presented below we detect light through an aperture corresponding to a 1.4 degree span, and therefore resulting in NA=0.0125. Transmission measurements using white light also allow characterization of the microcavity in areas away from the EL device.

\begin{figure}
\includegraphics[width=0.8\textwidth]{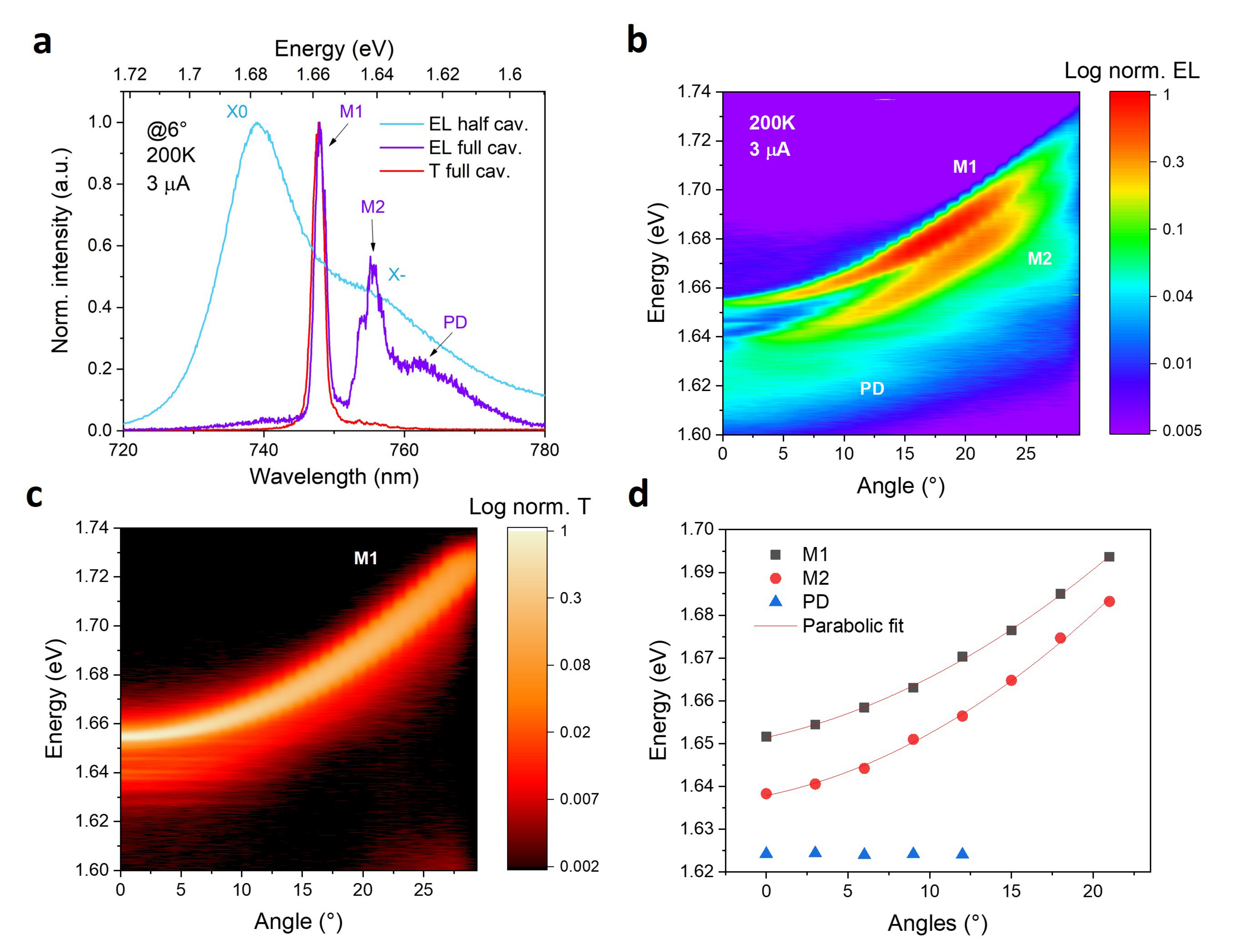}
\caption{\label{fig2} {\bf Optical properties of EL devices.} (a) Normalized EL spectra measured at $T=$200K for the current I=3$\mu$A before (blue) and after (purple) deposition of the top DBR. The EL in the full-cavity is measured for $\theta=6\degree$. A transmission spectrum of the microcavity measured outside the heterostructure at $T$=200K at 6$\degree$ is shown in red. $X^0$ and $X^-$ mark the peaks of the neutral exciton and trion in the half-cavity EL spectrum, respectively. M1 and M2 denote the cavity modes excited off and on the EL device, respectively (see text). PD is the broad feature corresponding to the photonic defects formed by the DBR distortions around the contacts. (b) Contour plot showing the angle-dependence of EL measured for $T$=200K and $I$=3$\mu$A in a full-cavity structure. (c)  Angle-dependent transmission spectra of the full-cavity structure measured outside the area of the EL device. (d) The peak positions of the main features observed in (a) and (b) can be fitted by two parabolas, corresponding to the dispersions of the cavity modes M1 (squares) and M2 (circles). No dependence of the peak energy on the angle is observed for the PD feature (triangles).}
\end{figure}

As demonstrated in Fig.\ref{fig1}(g) (right panel), where the current through the device versus the applied bias is plotted, the electrical functionality of the device is very similar in the full- and half-cavity configurations, owing to the protective layers of graphene and hBN, which most likely prevented the damage of WSe$_2$ during the top DBR deposition. Fig.\ref{fig2}(a) shows normalized EL spectra at temperature of 200K in both mirror configurations together with the transmission spectrum of the cavity measured outside the heterostructure. In the half-cavity device (blue line), the EL spectrum consists of two broad peaks corresponding to the neutral exciton $X^0$ (1.68 eV) and trion $X^-$ (1.64 eV) \cite{AroraNscale2015,Withers2015NanoLett}. EL spectrum for the full cavity measured at 6$\degree$ to normal incidence shows a more complex pattern (purple line). The dominant feature is a pronounced narrow peak at 1.658 meV labeled M1. This peak coincides with the microcavity mode measured in white light transmission (red line) through the cavity outside the heterostructure. Additional features are observed in the EL spectrum at lower energies of around 1.638 and 1.62 meV which can be assigned respectively to another cavity mode and a photonic defect (M2 and PD), as demonstrated in the following section. 

In order to confirm the origin of the spectral features of the full-cavity device we carried out angle-dependent EL measurements shown in Fig.\ref{fig2}(b). The dispersions of M1, from 1.65 eV to 1.73 eV, and M2, from 1.64 eV to 1.71 eV, are clearly visible in Fig.\ref{fig2}(b) where the collection angle $\theta$ is increased from 0$\degree$ to 30$\degree$. The peak positions of M1, M2 and PD peaks are plotted in Fig.\ref{fig2}(d). Both M1 and M2, follow the expected parabolic dispersion typical for a planar microcavity \cite{Kavokin2007}. 

We explain the occurrence of the two pronounced parabolic modes by the local change of the optical cavity length at the position of the EL device, leading to formation of M2. The perfect match of the M1 peak dispersion with the cavity mode dispersion outside the device measured in transmission (Fig.\ref{fig2}(c)), reveals the origin of this mode, which is excited because of a so-called 'walk-off' of the light emitted from the WSe$_2$. This is common in microcavities if the mirrors are not perfectly parallel \cite{Kavokin2007} or in the presence of scattering, which in this case is probably occurring at the flake edges and at the gold contacts. The thickness of the EL device in our case is $\approx$6.5 nm. The observed energy separation between M1 and M2 of 16 meV can be accounted for by the change in the cavity thickness if we assume an effective refractive index of 2.2 for the EL device, a reasonable average value between those of hBN and WSe$_2$. The linewidth of the principal cavity mode M1, measured both in transmission and in EL, is 1.3 nm, corresponding to a high Q-factor of 580. M2 is slightly broader than M1 showing a Q-factor of $\approx$250 when measured at 6$\degree$. This broadening is likely to be related to the presence of graphene and WSe$_2$ itself, whose absorption affects the Q-factor of the cavity \cite{ImamogluNatPhys2017}. The fact that M1 is excited outside the device is beneficial for its Q-factor, since it avoids the absorption losses due to the graphene electrodes and WSe$_2$.  Note that for large $\theta$, the parabolic dispersions shown in Fig.\ref{fig2}(b-d) for M1 and M2 are steeper, and therefore a larger energy range is measured within the same angular span of 1.4 degrees. This explains the broadening of the dispersive cavity modes at larger collection angles. 

The angle-resolved measurements in Fig.\ref{fig2}(b) also reveal that the PD peak position does not depend on the angle. This 'photonic defect' cavity mode possibly occurs due to the cavity distortions around the contacts observed in Fig.\ref{fig1}(d), which forms 'zero-dimensional' modes exhibiting no dispersion \cite{Kavokin2007,Gutbrod1999}. Photonic defects in the microcavity, present due to the discontinuity of DBR growth around the gold contacts, should result in quantized modes due to the strong lateral confinement. This is indeed observed in Fig. \ref{fig2}(d), where the confined PD mode is observed for collection angles up to 15 degrees and shows no dependence of its position on the angle of observation. The angular spread of the mode corresponds to a dimension in the momentum space which is obtained by the projection of the wave vector $k=\frac{2\pi}{\lambda}\text{sin}\theta$, where $\lambda$ is the wavelength of the mode and $\theta$ is the angle to which the cavity mode extends to. The Fourier transformation of the momentum space allows us to estimate a physical dimension of about $1.8\mu$m for the photonic defect which is of the order of magnitude of the lateral size of the gold contacts near the EL device.

\begin{figure}
\includegraphics[width=0.7\textwidth]{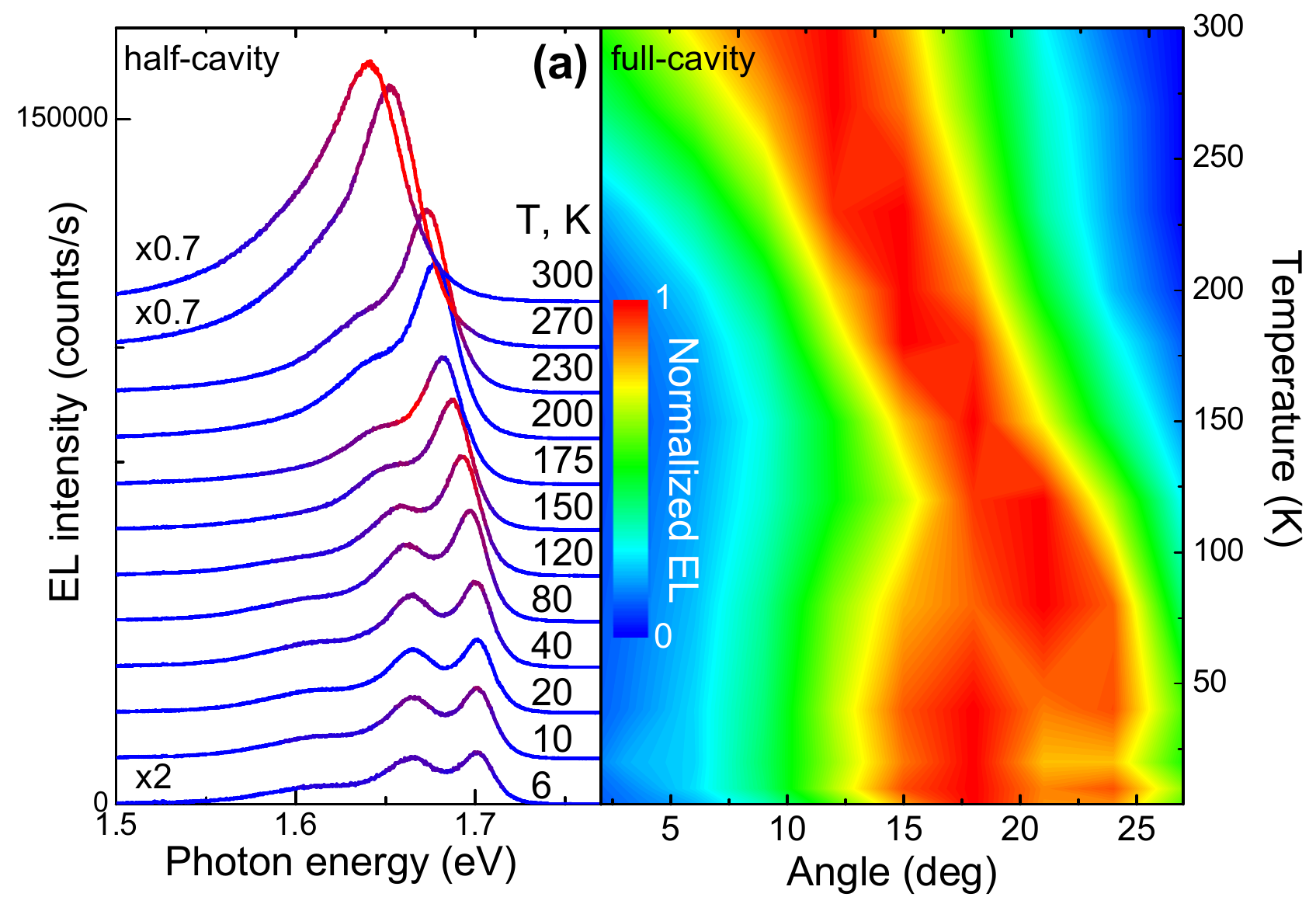}
\caption{\label{fig3} {\bf EL intensity as a function of the detection angle and temperature.} (a) The temperature dependence of the EL in the half-cavity configuration. A red-shift of the exciton energy and a large increase of intensity with increasing temperature is observed. All spectra are measured for the current 3$\mu$A passed through the device. (b) The angle dependence of the integrated EL intensity in a full-cavity configuration measured for different temperatures. The integrated EL intensity in each angle-dependence for a given temperature is normalized (see text).}
\end{figure}

From the comparison of the dispersions in Fig.\ref{fig2} with the half-cavity EL spectrum we conclude that for $\theta=0\degree$ both M1 and M2 have energies lower than that for $X^0$. The resonance condition between the cavity mode and the exciton or trion luminescence defines the angle at which the maximum intensity is reached in the microcavity. This resonant condition can be modified by changing the device temperature. As shown in Fig.\ref{fig3}(a) for the half-cavity device, EL red-shifts with increasing temperature following the decrease of the WSe$_2$ band-gap. At low $T$, two distinct EL peaks are observed corresponding to $X^0$ and $X^-$ at high and low energies, respectively. Above 150 K, the $X^-$ peak is only observed as a shoulder while the intensity of $X^0$ gradually increases in agreement with previous reports for WSe$_2$\cite{Withers2015NanoLett}. The angle- and temperature-dependence of the EL in the full-cavity device is shown in Fig.\ref{fig3}(b). For each reported temperature, the EL is measured as a function of the angle. Then the integrated EL intensity is calculated for each angle and is normalized by the maximum integrated EL value in the given angle scan. A horizontal slice of the normalized EL data in Fig.\ref{fig3}(b) corresponds to a full angle scan at a given temperature. The strongest EL is observed at the angle where the M1 mode is in resonance with the EL maxima of WSe$_2$. At low $T$ where both $X^0$ and $X^-$ peaks are strong, the enhancement of M1 intensity is observed in a broad angular range 15-25 degrees. The range where enhanced EL is observed narrows around 150 K, where the EL maximum of WSe$_2$ corresponding to $X^-$ is reduced in intensity, while the $X^0$ peak still remains relatively narrow. As both $X^0$ and $X^-$ become much broader for $T>250$K, the angular range with high cavity EL intensity increases again. Moreover, the angle where the cavity EL maximum is observed gradually moves to smaller values as the temperature is increased: at room temperature the integrated EL intensity reaches the maximum at around 13$\degree$.

\begin{figure}
\includegraphics[width=0.6\textwidth]{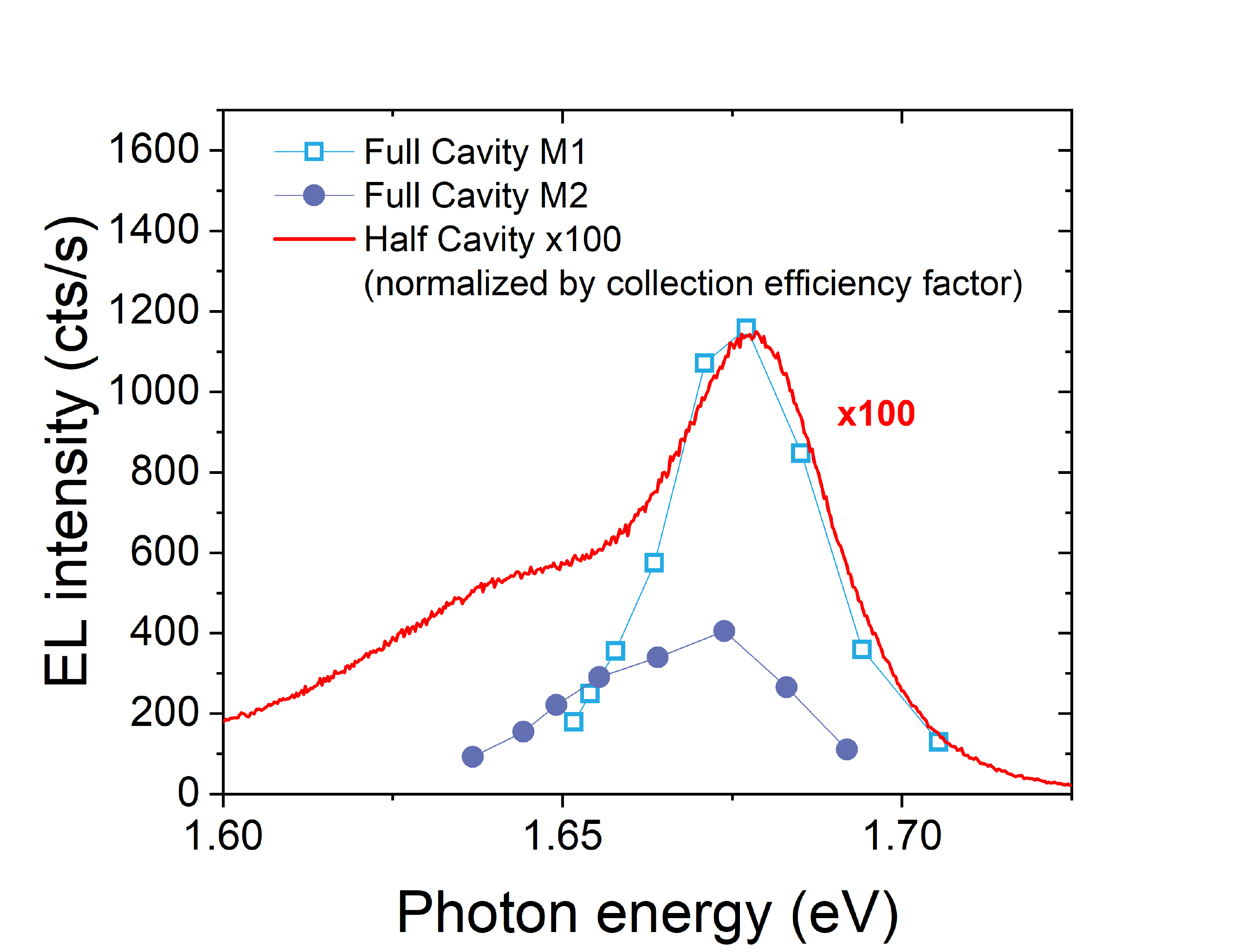}
\caption{\label{fig4} {\bf Peak EL intensity in half- and full-cavity devices.} EL spectrum measured for a half-cavity device at 200 K for I=3$\mu$A is shown in red, normalized by the collection efficiency factor calculated for the employed optical setups. The resulting signal intensity is multiplied by 100. Peak intensities for M1 and M2 mode spectra measured at different angles are shown with squares and circles, respectively.} 
\end{figure}

The micro-PL set-up used to characterize the half-cavity devices has a large NA of 0.55, and as a result its collection efficiency is $\approx$1900 times higher compared to the angle-resolved set-up, as follows from the ratio of NA$^2$ for each set-up. Fig.\ref{fig4} shows a comparison of the EL spectrum measured in a half-cavity configuration normalized by the collection efficiency factor (solid red line) with the peak intensities of modes M1 (squares) and M2 (circles) as they scanned through the emission angle, being in resonance with the X$_0$ peak at around 15$\degree$ and 18$\degree$, respectively. 

Taking into account that the cavity mode EL is collected through a low NA lens, then the results of Fig.\ref{fig4} show an enhancement of $\approx$100 and $\approx$35 of M1 and M2 EL, respectively, compared to the half-cavity EL. This observation constitutes a very significant narrowing of the angular width of the emission pattern in the full-cavity device as shown in Figs.\ref{fig3} and \ref{fig4} compared with the EL device placed on a DBR.

The weaker emission of M2 mode may occur because of the additional re-absorption that light in the M2 mode experiences due to the resonance with the exciton transition in the TMD. On the other hand, EL of the M1 mode occurring due to the 'walk-off' of the emitted light outside the area of the device does not suffer from such re-absorption processes. The shapes of the resonances for the two modes are also quite different, reflecting the additional suppression of the M2 mode in resonance with the exciton transition due to re-absorption and thus non-radiative losses due to the relatively low quantum efficiency of WSe$_2$. Such re-absorption processes are expected to be weak for the mode in resonance with X$^-$, which has lower oscillator strength \cite{Dufferwiel2015}.

\section{Conclusion}

In summary, we show for the first time an electroluminescent monolithic cavity based on a light emitting van der Waals heterostructure made from atomically thin materials. We find that the performance of the device is fully preserved after the top DBR deposition. The light emitted from WSe$_2$ is coupled into two narrow cavity modes arising from areas of the cavity containing the WSe$_2$ device and outside it, as well as to additional localized modes stemming from the photonic defects present in the structure. The bright cavity modes show very narrow linewidths leading to Q-factors of 580 and 250 respectively. The cavity also provides an increased EL directionality for the device, showing a peak intensity enhancement between 35 and 100 times at the angle of the maximum emission. We show that the EL re-absorption in the device may cause reduced intensity of emitted light, which could possibly be overcome by improving the quantum efficiency in WSe$_2$. The cavity EL can be spectrally and angularly tuned by changing the angle of collection of light or the operating temperature, respectively, which are additional degrees of freedom for the device tunability. Further improvements in fabrication, for example through using larger area materials, could improve the quality of the cavity mode. Future developments should see incorporation of several vertically stacked TMD monolayers separated by thin hBN layers, which should allow TMD-based vertical-cavity surface-emitting lasers (VCSELs) and electrically driven polariton devices.     \\

{\bf ACKNOWLEDGMENTS}

The authors thank the financial support of the European Graphene Flagship Projects under grant agreement 696656 and 785219, EC Project 2D-SIPC, and EPSRC grants EP/P026850/1 and EP/N031776/1. A.I.T. acknowledges support from the European Union's Horizon 2020 research and innovation programme under  ITN Spin-NANO Marie Sklodowska-Curie grant agreement no. 676108 and ITN S$^3$NANO Marie Sklodowska-Curie grant agreement no. 289795. O.D.P.Z was supported by CONACYT-Mexico. K.S.N. thanks financial support from the Royal Society, EPSRC, US Army Research Office (W911NF-16-1-0279) and ERC Grant Hetero2D. K.W. and T.T. acknowledge support from the Elemental Strategy Initiative conducted by the MEXT, Japan and the CREST (JPMJCR15F3), JST. S.J.H., E.P. and A.P.R. acknowledge support from the EPSRC, the Defense Threat Reduction Agency HDTRA1-12-1-0013  and ERC Starter Grant EvoluTEM (715502). \\ 

{\bf AUTHOR CONTRIBUTIONS}

S. S. and O. D.-P. Z. carried out optical investigations with contributions from  A. G., F. W., T. G. and R. C. S. TMD heterostructures were fabricated by F. W. C. C. fabricated DBRs. STEM images were taken by A. P. R., E. P. and S. J. H. K. W. and T. T. grew the hBN crystals. S. S., O. D.-P. Z., A. G., P. M. W., A. I. T. analyzed the data and wrote the manuscript with contributions from all co-authors. P. M. W. carried out numerical simulations. S. J. H., K. S. N., D. N. K. and A. I. T. provided management of various aspects of the project, and contributed to the analysis and interpretation of the data and writing of the manuscript. A. I. T. conceived and oversaw the whole project. 

\newpage

\section{Supporting Information}

{\bf MICROCAVITY DESIGN}

\begin{figure}[H]
	\centering
	\includegraphics[width=.8\textwidth]{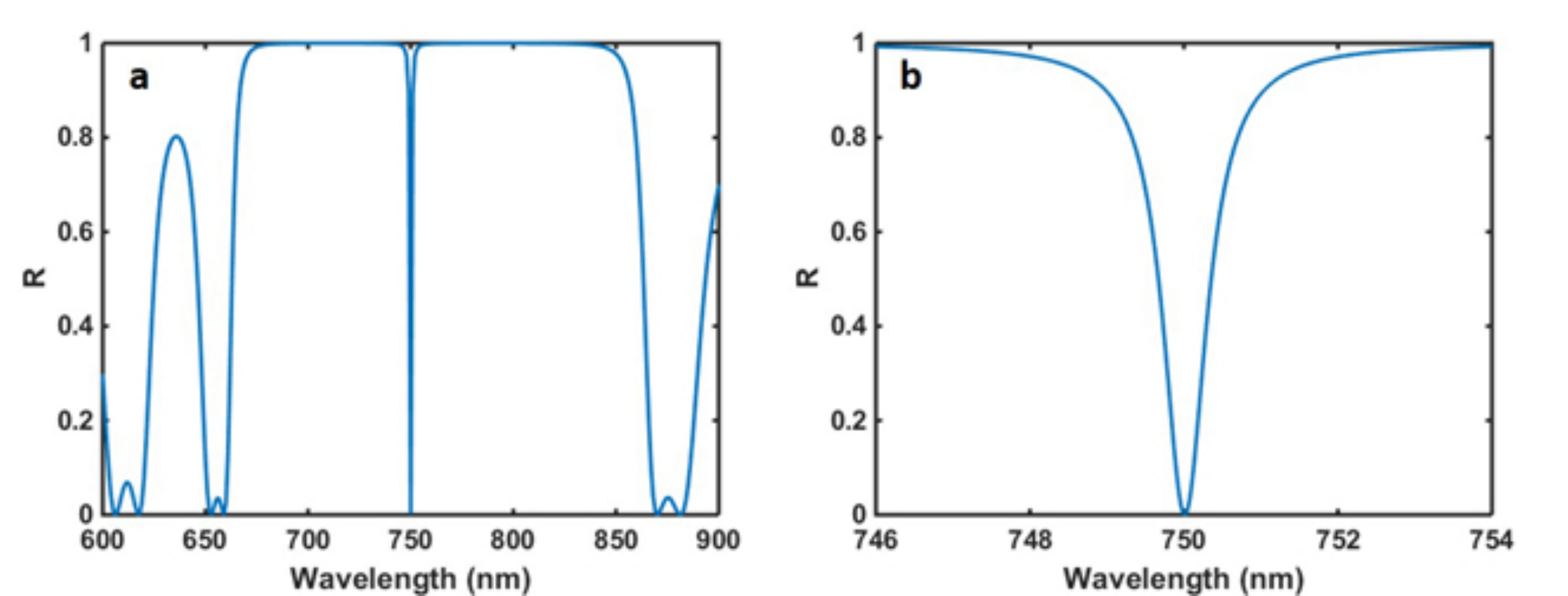}
	\caption{\label{design} {\bf Numerical simulations of the cavity reflectivity.} Transfer-matrix simulations have been used in order to design the microcavity structure, where 10 pairs of SiO$_2$/Nb$_2$O$_5$ have been used for the distributed Bragg reflectors (DBR) and an additional half-lambda SiO$_2$ layer serves as a microcavity. We used Nb$_2$O$_5$ and SiO$_2$ layers with thicknesses of 127 and 95 nm and refractive indexes of 1.45 and 2.09, respectively. (a) Reflectivity of a full-cavity device. The simulations show a cavity mode at 750 nm and a stop-band of 200 nm wide, for which a reflectivity of more than 99\% is calculated. (b) Zoom-in around the cavity mode wavelength. The simulated Q-factor of the microcavity is 1100, which is higher than the 580 observed in the main text due to fabrication imperfections.}
\end{figure}

\newpage

{\bf HAADF AND EDXS OF THE GOLD CONTACT}

\begin{figure}[H]
	\centering
	\includegraphics[width=.6\textwidth]{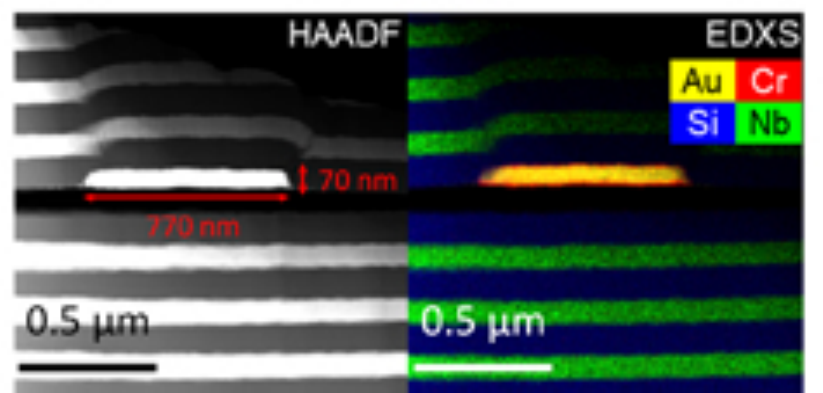}
	\caption{\label{HAADF} {\bf HAADF and EDXS} STEM High Angle Annular Dark Field (HAADF) image of the mirror pairs surrounding one gold contact. A map of chemical composition acquired by Energy Dispersive X-ray Spectroscopy (EDXS) is shown on the right. SiO$_2$ and Nb$_2$O$_5$ mirror pairs, shown blue and green respectively, surround the gold contact, yellow, and cavity, shown in  black. Scale bar 500 nm.}
\end{figure}

{\bf EELS MAP OF THE HETEROSTRUCTURE}

\begin{figure}[H]
	\centering
	\includegraphics[width=.6\textwidth]{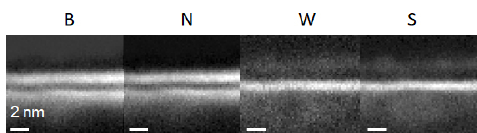}
	\caption{\label{EELS} {\bf EELS of the heterostructure} Electron energy loss spectroscopy (EELS) mapping of hBN encapsulated WSe$_2$ monolayer for different elements.}
\end{figure}

\newpage

{\bf EFFECT OF THERMAL CYCLES ON THE STABILITY OF THE DEVICE}

\begin{figure}[H]
	\centering
	\includegraphics[width=.6\textwidth]{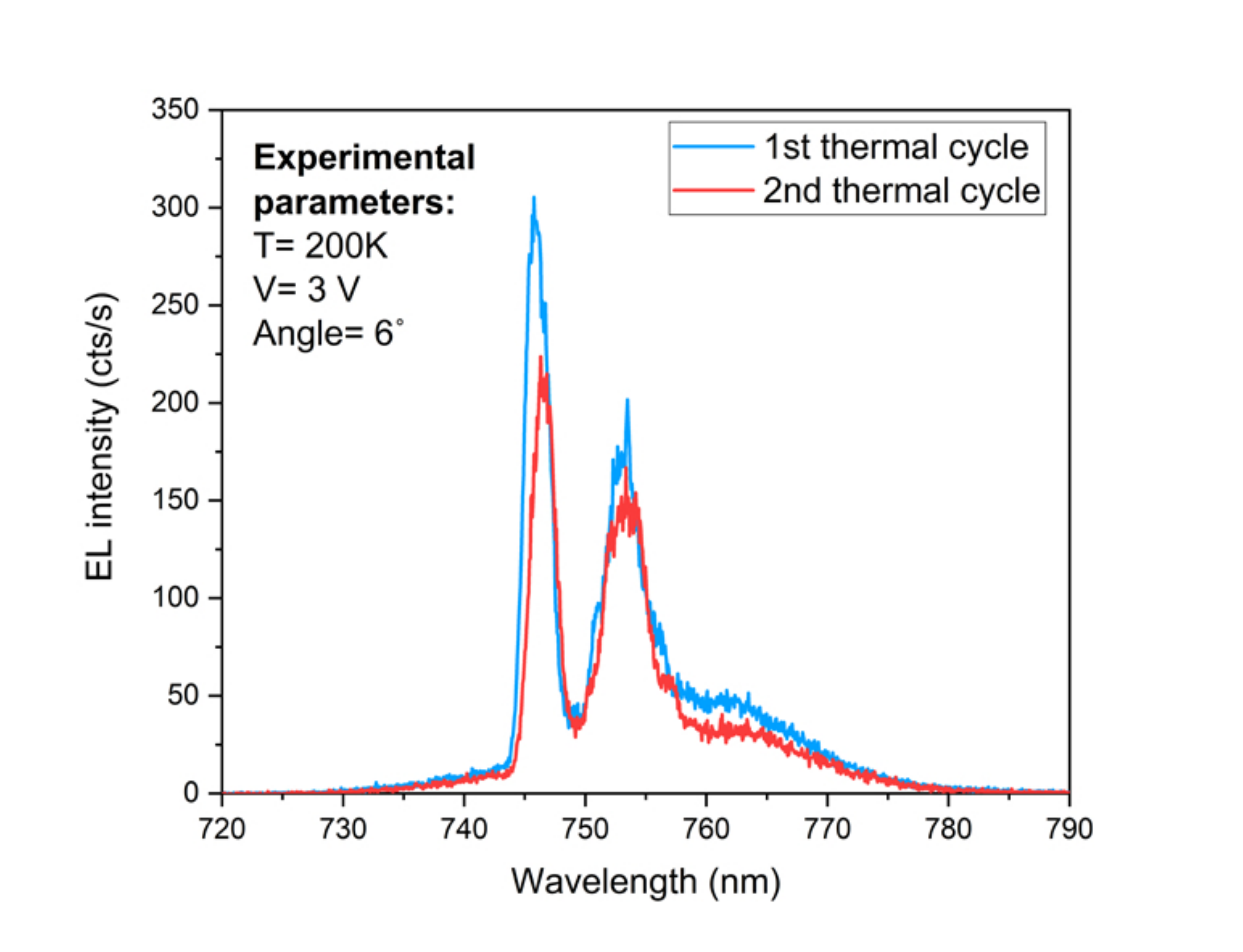}
	\caption{\label{EELS} {\bf Stability test on the EL device} EL spectra of the monolithic cavity device recorded during the first (blue line) and second (red line) thermal cycles for the same experimental conditions (Temperature: 200K, Voltage: 3V, Angle: 6$\degree$). We found that the main source of degradation in our measurements, although still relatively weak, was the effect of thermal cycling. Indeed, the EL intensity decreased by a factor of about 0.7 in the end of the second thermal cycle. A slight emission red shift of a few nanometres has also occurred (which could be caused by a slightly different temperature of the device).}
\end{figure}


\newpage

\begin{thebibliography}{28}

\bibitem{Dean2010} {Dean, C. R. {\it et al.} Boron nitride substrates for high-quality graphene electronics, {\it Nature Nanotech.} {\bf 5} 722-726 (2010)}

\bibitem{NovoselovPNAS2005} {Novoselov, K. S. {\it et al.} Two-dimensional atomic crystals, {\it PNAS} {\bf 102} 10451-10453 (2005)}

\bibitem{Geim2013} {Geim, A. K. \& Grigorieva, I. V. Van der Waals heterostructures, {\it Nature} {\bf 499} 419-425 (2013)}

\bibitem{Britnell2013} {Britnell, L. {\it et al.} Strong Light-Matter Interactions Thin Films, {\it Science} {\bf 340} 1311-1314 (2013)}

\bibitem{WangNatNano2012} {Wang, Q. H., Kalantar-Zadeh, K., Kis, A., Coleman, J. N., \& Strano, M. S. {Electronics and optoelectronics of two-dimensional transition metal dichalcogenides}, {\it Nature Nanotech.} {\bf 7} 699-712 (2012)}  
   
\bibitem{XuNatPhys2014} {Xu, X., Yeo, W., Xiao, D. \& Heinz, T. F. {Spin and pseudospins in layered transition metal dichalcogenides}, {\it Nature Physics} {\bf 10} 343-350 (2014)}  

\bibitem{MakPRL2010} {Mak, K., Lee, C., Hone, J., Shan, J., \& Heinz, T. {Atomically Thin MoS$_2$: A New Direct-Gap Semiconductor}, {\it Phys. Rev. Lett.} {\bf 105} 2-5 (2010)}

\bibitem{Ugeda2014} {Ugeda, M. M. {\it et al.} {Giant bandgap renormalization and excitonic effects in a monolayer transition metal dichalcogenide semiconductor}, {\it Nature Mater.} {\bf 13} 1091-1095 (2014)}  

\bibitem{Berkelbach2013} {Berkelbach, T. C., Hybertsen, M. S. \& Reichman, D. R. {Theory of neutral and charged excitons in monolayer transition metal dichalcogenides}, {\it Phys. Rev. B} {\bf 88} 045318 (2013)}  

\bibitem{Mak2016} {Mak, K., F., \& Shan, J. {Photonics and optoelectronics of 2D semiconductor transition metal dichalcogenides}, {\it Nat. Phot.} {\bf 10} 216-226 (2016)}

\bibitem{Withers2015} {Withers, F. {\it et al.} {Light-emitting diodes by band-structure engineering in van der Waals heterostructures}, {\it Nature Mater.} {\bf 14} 301-306 (2015)}

\bibitem{Withers2015NanoLett} {Withers, F. {\it et al.} {WSe$_2$ Light-Emitting Tunneling Transistors with Enhanced Brightness at Room Temperature}, {\it Nano Lett.} {\bf 15} 8223-8228 (2015)}

\bibitem{Pospischil2014} {Pospischil, A. {\it et al.} {Solar-energy conversion and light emission in an atomic monolayer p-n diode}, {\it Nature Nanotech.} {\bf 9} 257-261 (2014)}

\bibitem{Ross2014} {Ross, J. S. {\it et al.} {Electrically tunable excitonic light-emitting diodes based on monolayer WSe$_2$ p-n junctions}, {\it Nature Nanotech.} {\bf 9} 268-272 (2014)}     


\bibitem{EdaAdvMat2018} {Wang, J., Verzhbitskiy, I., \& Eda, G. {Electroluminescent Devices Based on 2D Semiconducting Transition Metal Dichalcogenides}, {\it Adv. Mat.} {\bf 30} 1802687, 1-14 (2018)}

\bibitem{SchubertScience1994} {Schubert, E. F., {\it et al.} {Highly efficient light-emitting diodes with microcavities}, {\it Science} {\bf 265.5174} 943-945 (1994)}

\bibitem{Soda1979} {Soda, H. {\it et al.} {GaInAsP/InP Surface Emitting Injection Lasers}, {\it Jap. Journal of Appl. Phys.} {\bf 18} 2329-2330 (1979)}

\bibitem{Koyama1989} {Koyama, F. {\it et al.} {Room temperature cw operation of GaAs vertical cavity surface emitting laser}, {\it Appl. Phys. Lett.} {\bf 55} 221-222 (1989)}

\bibitem{Michalzik2012} {Michalzik, R. {VCSELs: fundamentals, technology and applications of vertical-cavity surface-emitting lasers}, {\it Springer} Vol. 166 (2012)}


\bibitem{Wu2014} {Wu, S. {\it et al.} {Control of two-dimensional excitonic light emission via photonic crystal}, {\it 2D Mater.} {\bf 1} 011001 (2014)}

\bibitem{Gan2013} {Xuetao, G. {\it et al.} {Controlling the spontaneous emission rate of monolayer MoS$_2$ in a photonic crystal nanocavity}, {\it App. Phys. Lett.} {\bf 103} 181119 (2013)}   

\bibitem{Schwarz2014} {Schwarz, S. {\it et al.} {Two-Dimensional Metal-Chalcogenide Films in Tunable Optical Microcavities}, {\it Nano Lett.} {\bf 14} 7003-7008 (2014)}

\bibitem{Wu2015} {Wu, S. {\it et al.} {Ultra-Low Threshold Monolayer Semiconductor Nanocavity Lasers}, {\it Nature} {\bf 520} 520 (2015)}

\bibitem{Salehzadeh2015} {Salehzadeh, O. {\it et al.} {Optically Pumped Two-Dimensional MoS$_2$ Lasers Operating at Room-Temperature}, {\it Nano Lett.} {\bf 15(8)} 5302-5306 (2015)}

\bibitem{Ye2015} {Ye, Y. {\it et al.} {Monolayer excitonic laser}, {\it Nat. Phot.} {\bf 9.11} 733 (2015)} 

\bibitem{Shang2017} {Shang, J., {\it et al.} {Room-temperature 2D semiconductor activated vertical-cavity surface-emitting lasers.}, {\it Nature communications} {\bf 8.1} 543 (2017)}

\bibitem{Liu2015} {Liu, X. {\it et al.} {Strong light-matter coupling in two-dimensional atomic crystals}, {\it Nature Phot.} {\bf 9} 30-34 (2015)}

\bibitem{Dufferwiel2015} {Dufferwiel, S. {\it et al.} {Exciton-polaritons in van der Waals heterostructures embedded in tunable microcavities}, {\it Nature Communications} {\bf 6}, 8579 (2015)}

\bibitem{Lundt2016} Lundt, N. {\it et al.} Room-temperature Tamm-plasmon exciton-polaritons with a WSe$_2$ monolayer. {\it Nature Communication} {\bf 7}, 13328 (2016).

\bibitem{Sidler2017} Sidler, M. {\it et al.} Fermi polaron-polaritons in charge-tunable atomically thin
semiconductors. {\it Nature Physics} {\bf 13}, 255 (2017).

\bibitem{SchneiderReview2018} Schneider, C., Glazov, M. M., Korn, T.,  Hoefling, S., Urbaszek, B. Two-dimensional semiconductors in the regime of strong light-matter coupling. {\it Nature Communications} {\bf 9} 2695 (2018). 

\bibitem{Liu2016} Liu, C.H., Clark, G., Fryett, T., Wu, S., Zheng, J., Hatami, F., Xu, X. and Majumdar, A.. Nanocavity integrated van der Waals heterostructure light-emitting tunneling diode. {\it Nano letters}, 17(1), pp.200-205 (2016). 

\bibitem{Gu2019} Gu, J., Chakraborty, B., Khatoniar, M. and Menon, V.M.. A Room Temperature Polariton Light-Emitting Diode Based on Monolayer WS2. arXiv preprint arXiv:1905.12227, 2019.

\bibitem{AroraNscale2015} {Arora, A. {\it{et al,}} {Excitonic resonances in thin films of WSe 2: from monolayer to bulk material.}, {\it Nanoscale} {\bf 7.23} 10421-10429 (2015)} 


\bibitem{Kavokin2007} {Kavokin, A. V., Baumberg, J. J., Malpuech, G. \& Laussy, F. P. {Microcavities}, {\it Oxford University Press} Series on Semiconductor Science and Technology (2007)}

\bibitem{Gutbrod1999} {Gutbrod, T. {\it et al.} Angle dependence of the spontaneous emission from confined optical modes in photonic dots, {\it Phys. Rev. B} {\bf 59} 2223-2229 (1999)}

\bibitem{ImamogluNatPhys2017} {Sidler, M. {\it et al.} Fermi polaron-polaritons in charge-tunable atomically thin semiconductors, {\it Nature Physics} {\bf 13.3} 255 (2017)}

\bibitem{YanAPL2014} {Yan, T. {\it et al.} Photoluminescence properties and exciton dynamics in monolayer WSe$_2$, {\it Appl. Phys. Lett.} {\bf 105.10} 101901 (2014)}

\bibitem{KretininNanoLett2014} {Kretinin, A. V. {\it{et al,}} {Electronic Properties of Graphene Encapsulated with Different Two-Dimensional Atomic Crystals}, {\it Nano Lett.} {\bf 14} 3270-3276 (2014)} 


\end{thebibliography}
\end{document}